# Relationship of temperature changes in the mesopause region with the climate changes at the surface from observations in 1960-2024

Mokhov I.I.[1,2], Fomina I.A.[2], Perminov V.I.[1]

[1]A.M. Obukhov Institute of Atmospheric Physics RAS

[2]Lomonosov Moscow State University

mokhov@ifaran.ru

**Abstract**

The results of an analysis of temperature variations in the mesopause region based on long-term measurements of hydroxyl airglow at the Zvenigorod Scientific Station of the A.M. Obukhov Institute of Atmospheric Physics RAS (ZSS IAP RAS) in 1960–2024 in comparison with variations of surface temperature characterizing global-scale climate changes are presented. Along with temperature variations in the mesopause region, two versions of temperature variations in the mesopause region, normalized to the same level of solar activity, were analyzed. Quantitative estimates of a strong decrease in temperature in the mesopause region over the past decades in winter against the background of a global increase in surface temperature have been obtained. It was noted that significant coherence of long-term variations for temperature in the mesopause region with the surface temperature in the Northern Hemisphere with the use of cross-wavelet analysis, what was not previously evident in data for a shorter time interval. The possibility of such coherence was predicted in (Mokhov et al., 2017) under the continuation of global warming based on the results of model simulations for the 20−21 centuries, taking into account anthropogenic forcing. It was not previously manifested from observational data for a shorter time interval. Along with long-term trends, features of a sharp decrease in temperature in the mesopause region in the 1970s with its synchronicity with the known shift in surface climate regimes associated with El Niño events were analyzed. The results of cross-wavelet analysis using data obtained at the ZSS IAP RAS for the time interval 1960-2024 indicate a more significant connection between temperature variations in the mesopause region and El Niño indices in recent decades.

**INTRODUCTION**

The problem of climate change is one of the key modern problems. To adequately understand ongoing climate change, it is necessary to determine the relative role of natural and anthropogenic factors. The analysis of vertical profiles of temperature trends in the atmosphere is of particular importance. Under global warming at the surface and in the troposphere in recent decades, a general cooling of the stratosphere and mesosphere has been revealed. The general cooling of the middle and upper atmosphere under general warming at the surface, noted from the results of various measurements, is an important indicator of the potential role of anthropogenic mechanisms of global climate change. The results of model simulations indicate significant differences in temperature changes in different atmospheric layers, in particular in the stratosphere and mesosphere, depending on the type of external forcing on the climate system: natural (including solar and volcanic activity) and anthropogenic (including changes in the content of greenhouse gases and aerosol in the atmosphere) ones [1−19]. According to model simulations, under the increase in the content of greenhouse gases (carbon dioxide and others) in the atmosphere against the background of general warming at the surface and in the troposphere, a cooling of the atmospheric layers above the tropopause is expected (Fig. 1). At the same time, due to natural variations in insolation, a general synphase of temperature variations in the lower and middle atmosphere appears (see, for example, [8,20,21]).

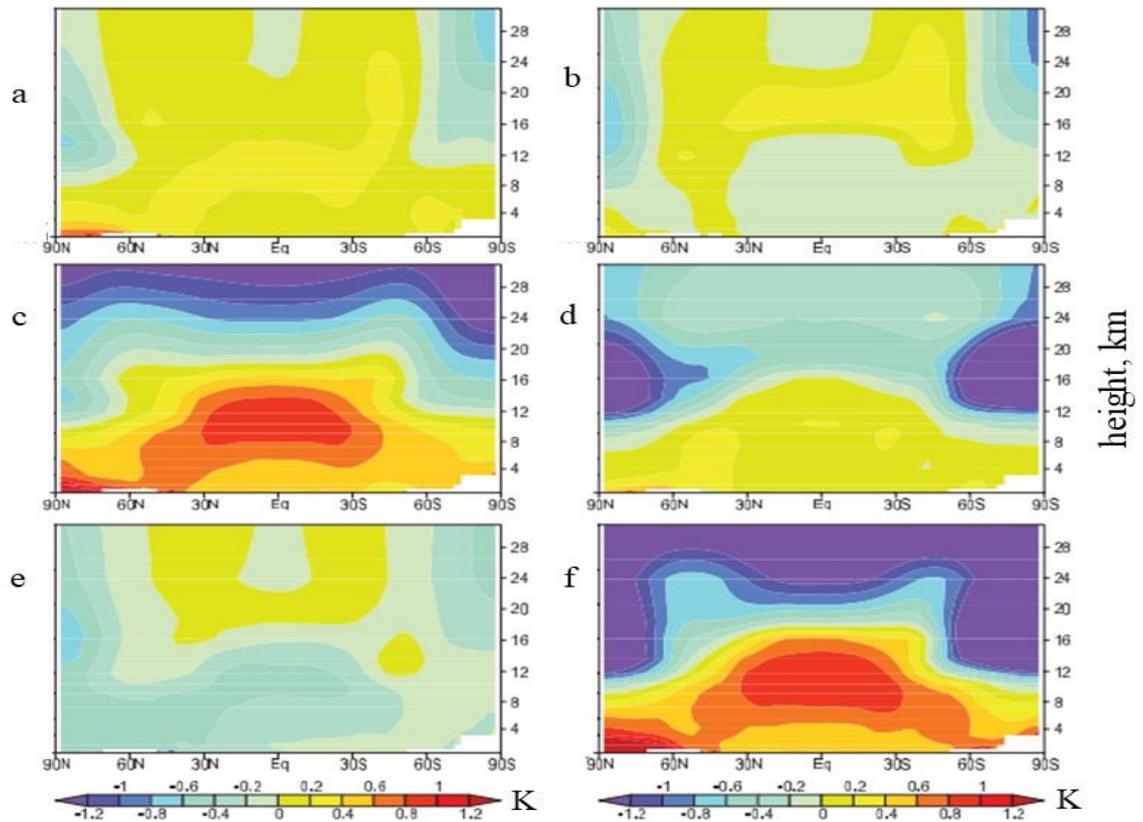

Fig. 1. Model estimates of latitude-altitude distributions for changes in atmospheric temperature [K/century] from 1890 to 1999, taking into account changes in (a) solar forcing, (b) volcanic activity, (c) greenhouse gases, (d) tropospheric and stratospheric ozone, (e) radiative forcing of sulfate aerosols and (f) total forcing [20] (see also [8, 21]).

In the Earth's climate system, a special temperature regime is distinguished in the mesopause region, i.e. at the boundary between the mesosphere and thermosphere (at altitudes of 80–100 km) with the lowest temperature values. It is no coincidence that a global network of stations has been created that monitors changes in the mesopause region, so called NDMC (Network for the Detection of Mesopause Change, http://wdc.dlr.de/ndmc/). The key argument for the creation of the international NDMC project is the possibility of early detection of global changes in the temperature regime of the atmosphere and its composition from measurements in the upper atmosphere, since the reaction of rarefied layers of the upper atmosphere, in particular in the mesopause region, is more sensitive to both the effects of solar radiation and other forcings, including anthropogenic ones, compared with the response for the lower atmosphere. It should be noted that the lower layers of the atmosphere are significantly influenced by various dynamic processes that arise when interacting with inhomogeneities on the Earth's surface [19]. Within the framework of the NDMC system, the longest time series of data for atmospheric temperature in the mesopause region is based on measurements at the Zvenigorod Scientific Station of the A.M. Obukhov Institute of Atmospheric Physics RAS (ZSS IAP RAS) since the 1950s [7].

**DATA AND METHODS OF ANALYSIS**

The data for the temperature $T_{ms}$ in the mesopause region by measurements at the Zvenigorod Scientific Station (56° N, 37° E) of the A.M. Obukhov Institute of Atmospheric Physics RAS in 1959–2024 were analyzed. The values of $T_{ms}$ were determined (with an accuracy of 2K) from the spectrometric measurements of hydroxyl airglow with the emitting maximum at an altitude of about 87 km [7] (see also [14, 16, 19]).

In 2015, a series of spectrophotometric measurements of temperature around the mesopause at the ZSS IAP RAS together with measurements in Moscow region of vertical distributions of temperature and ozone content in the atmosphere using a multifunctional high-altitude sensing lidar, were carried out [22]. During lidar soundings, altitude distributions of atmospheric temperature were determined up to an altitude of about 100 km. Along with this, temperature profiles obtained from lidar measurements were compared with satellite data (AURA, TIMED/SABER) and the CIRA model. The obtained comparison results indicate a fairly good agreement between the various data, in particular for the temperature in the mesopause region.

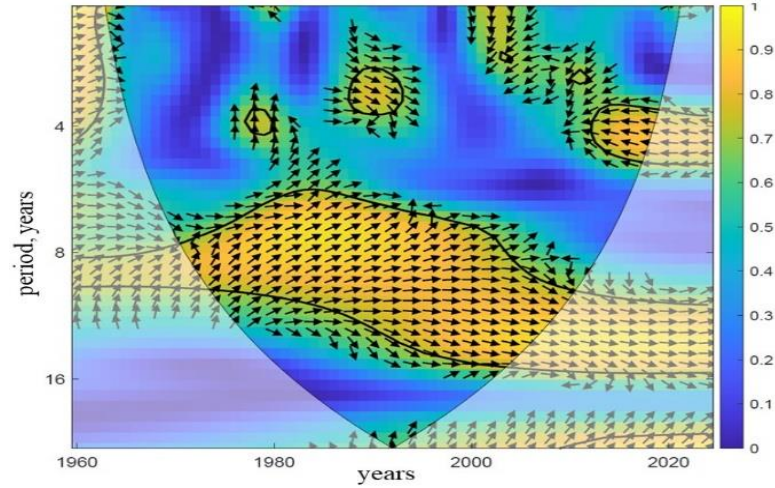

Fig. 2. Cross-wavelet coherence of winter temperature variations in the mesopause region $T_{ms}$ from observations for the period 1960–2024 with solar activity index F10.7. Solid lines delineate areas of edge effects, and thick lines delineate areas of significant coherence (with non-zero coherence at the p = 0.05 significance level). The arrows characterize the phase shift: arrows to the right mean in-phase, arrows to the left mean anti-phase.

We analyzed also the variations in the temperature in the mesopause region, $T_m$, obtained by the reduction of the $T_{ms}$ values to the constant level of solar activity characterized by the index of the flux density of solar radiation $F$10.7 (http://www.wdcb.ru/stp/data/solar.act/flux10.7/daily/) (Fig. 2). In this case

$$T_{ms} = T_m + \delta T_{ms},$$

where the contribution of solar radiation flux variations to temperature variations in the mesopause region $\delta T_{ms}$ was estimated similar to [7]. Temperature values in the mesopause region $T_m(1)$ were analyzed, taking into account

$$\delta T_{ms} = 25 \log [F10.7 \, (t - 0.42)/150], \qquad (1)$$

where $t$ is time in years. The parameterization (1) was obtained based on observations without taking into account seasonality (see [7]). It was recently noted that winter and summer temperatures in the mesopause region have different responses to changes in solar activity [23]. It should be also noted, that for the winter season the response is linear and without time lag:

$$\delta T_{ms} / \delta F10.7 = 0.054 \text{ K/sfu},  \quad (2)$$

sfu is the unit of measurement of F10.7 solar flux. Along with $T_m(1)$, the temperature values in the mesopause region $T_m(2)$ were analyzed, reduced to the level of solar activity F10.7=130 taking into account (2).

It is necessary to note that due to gaps in measurements (associated with cloudiness) in the time interval 1960−2024 the winter mean (December to February) temperature values in the mesopause region were analyzed using approximations for annual time intervals (from July 1 of the previous year to June 30 of the next year), taking into account annual and semi-annual harmonics.

Along with estimates of the temperature trends in the mesopause region $dT_m(1)/dt$, $dT_m(2)/dt$ and $dT_{ms}(1)/dt$, the sensitivity parameters of temperature in the mesopause region to changes in the NH surface temperature $T_{NHs}$ were determined using the corresponding linear regressions: $dT_{ms}/dT_{NHs}$, $dT_m(1)/dT_{NHs}$, $dT_m(2)/dT_{NHs}$. We used monthly-mean values of the NH surface temperature anomalies (and for the Earth as a whole) relative to its base conditions for the period 1961–1990 (http://www.cru.uea.ac.uk/). For winter 2023−2024 the $T_{NHs}$ value was determined by the average value for December 2023 and January 2024.

The cross-wavelet analysis [24] was also used, in particular, the local coherence of temperature in the mesopause region with surface temperature, solar activity indices and El Niño phenomena indices, including the Nino3, Nino3.4, Nino4 indices (https://psl.noaa.gov/data/climateindices/).

The results of analysis of observational data were compared with results of model simulations for the 20th–21st centuries [19]. Figure 2 characterizes the cross-wavelet coherence of winter variations in hemispheric surface temperature $T_{NHs}$ and temperature in the mesopause region (at the level of 0.003 hPa) at latitude 56°N from model simulations taking into account anthropogenic forcings [19].

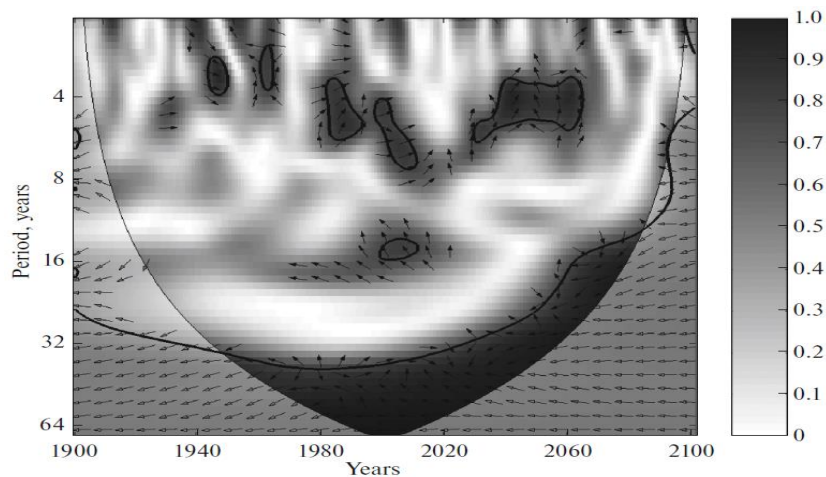

Fig. 3. Cross-wavelet coherence of winter variations in hemispheric surface temperature $T_{NHs}$ and temperature in the mesopause region (at the level of 0.003 hPa) at latitude 56° N from model simulations taking into account anthropogenic forcings [19]. Solid lines delineate areas of edge effects, and thick lines delineate areas of significant coherence (with non-zero coherence at the $p$ = 0.05 significance level). The arrows characterize the phase shift: arrows to the right mean in-phase, arrows to the left mean anti-phase.

According to Fig. 3 there is a shift since the end of the 20th century in the boundary of significant coherence (limited by the thick curve) towards shorter-period variations, namely to periods of about 20 years in the second half of the 21st century, and by the end of the century to periods of less than 10 years. According to model estimates, under continuation of global warming at the surface (and in the troposphere) with cooling of the strato-mesosphere, we have to expect in the next decades the manifestation of significant long-term coherence of temperature variations at the level of the mesopause and $T_{\text{NHs}}$ from the analysis of observational data.

**RESULTS**

Figure 4 shows interannual variations of $T_{\text{ms}}$ (a) for the winter season (from December to February) by measurements at the ZSS IAP RAS in 1960–2024 (more precisely, from December, 1959), as well as the corresponding variations of $T_{\text{m}}(1)$ (b) and $T_{\text{m}}(2)$ (c), in which variations of $\delta T_{\text{ms}}$ associated with time variations of the solar radiation flux $I_{\text{s}}(t)$, characterized by the F10.7 index, were filtered (see [7]). This was done by reducing the measured temperature values $T_{\text{ms}}$ (1960–2024) to the same level of solar activity in accordance with (1) and (2).

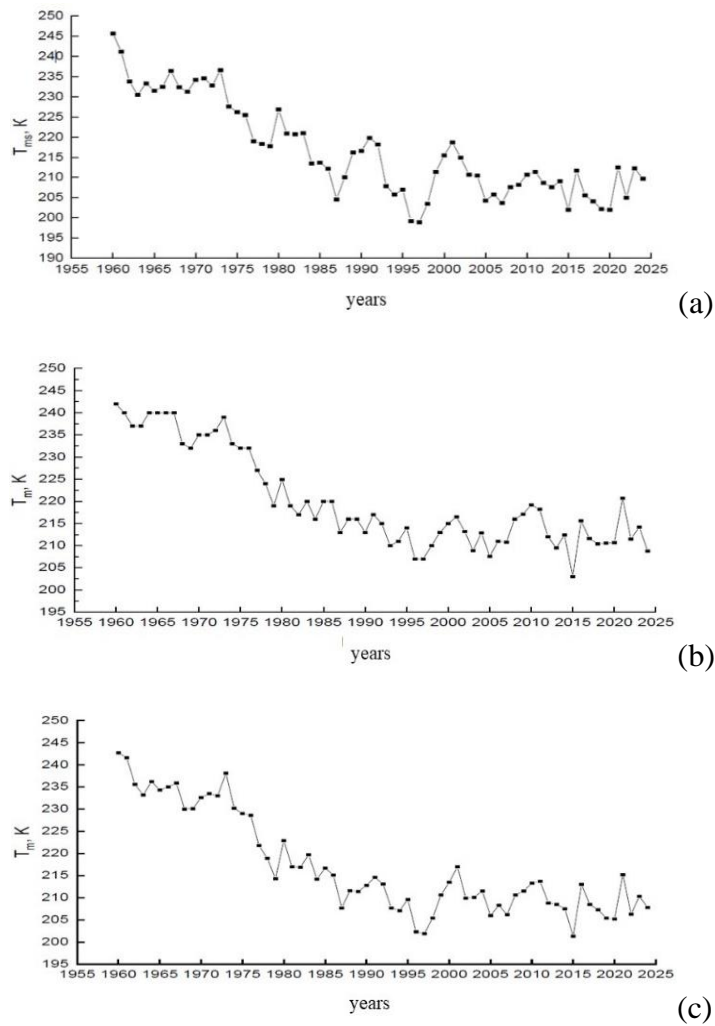

Fig. 4. Interannual variations in winter temperature in the mesopause region $T_{\text{ms}}$ (a), $T_{\text{m}}(1)$ (b) and $T_{\text{m}}(2)$ (c) from measurements at the ZSS IAP RAS in 1960–2024.

According to Fig. 4, the decrease in winter temperature at the mesopause level over the last 65 years has been about three tens of degrees. Average estimates of the linear trend $dT_m(1)/dt$ and $dT_m(2)/dt$ (at $I_s$ = const) for the entire analyzed period of measurements (1960–2024) about –5 K/(10 years) with a statistical significance of more than 99% (Table 1) are close to the value of the corresponding trend $dT_{ms}/dt$ taking into account variations in the solar radiation flux $I_s(t)$. Measurements at the ZSS IAP RAS revealed for 65 winters a strong nonlinearity of temperature changes in the mesopause region, with a tendency for a general slowdown in the decrease in recent decades. At the same time, for the first half of this time interval (1960-1992), statistically significant trends of about −10 K/(10 years) were obtained: $dT_m(1)/dt = -9.6$ K/(10 years), $dT_m(2)/dt = -9.9$ K/(10 years), and also $dT_{ms}(1)/dt = -8.9$ K/(10 years). For the second half (1992-2024), trend estimates are statistically insignificant (Table 1). A sharp decrease in temperature in the mesopause region was noted in the 1970s.

Table 1. Trend estimates $dT_{ms}/dt$ [K/(10 years)] (a), $dT_m(1)/dt$ [K/(10 years)] (b), $dT_m(2)/dt$ [K/(10 years)] (c), as well as the corresponding parameters of the relationship $dT_{ms}/dT_{NHs}$, $dT_m(1)/dT_{NHs}$, $dT_m(2)/dT_{NHs}$ for the entire analyzed time interval 1960−2024 and for two sub-intervals: 1960-1992 and 1992−2024 The standard deviations (SD) are indicated in parentheses. The most statistically significant estimates are highlighted.

(a)

| Parameters | 1960−2024 | 1960−1992 | 1992−2024 |
|---|---|---|---|
| $dT_{ms}/dt$ [K/(10 years)] | **−5.2(±0.4)** | **−8.9(±0.9)** | −0.5(±0.9) |
| $dT_{ms}/dT_{NHs}$ | **−15.8(±2.1)** | −14.3(±6.2) | −1.3(±2.4) |

(b)

| Parameters | 1960−2024 | 1960−1992 | 1992−2024 |
|---|---|---|---|
| $dT_m(1)/dt$ [K/(10 years)] | **−4.8(±0.4)** | **−9.6(±0.6)** | 0.3(±0.7) |
| $dT_m(1)/dT_{NHs}$ | **−15.2(±1.9)** | **−21.1(±5.7)** | −0.3(±1.9) |

(c)

| Parameters | 1960−2024 | 1960−1992 | 1992−2024 |
|---|---|---|---|
| $dT_m(2)/dt$ [K/(10 years)] | **−5.0(±0.4)** | **−9.9(±0.7)** | −0.1(±0.7) |
| $dT_m(2)/dT_{NHs}$ | **−15.5(±2.0)** | **−19.1(±6.2)** | −0.7(±1.9) |

In a more detailed analysis for four 16-year subintervals, statistically significant trend estimates were obtained only for the first two subintervals: $dT_m(1)/dt = -4.9$ K/(10 years), $dT_m(2)/dt = -5.8$ K/(10 years), $dT_{ms}(1)/dt = -6.6$ K/(10 years) for 1960-1976 and $dT_m(1)/dt = -8.2$ K/(10 years), $dT_m(2)/dt = -7.5$ K/(10 years), $dT_{ms}(1)/dt = -5.2$ K/(10 years) for 1976−1992 (Table 2).

Table 2. Trend estimates d$T_{ms}$/dt [K/(10 years)] (a) d$T_m$(1)/dt [K/(10 years)] (b), d$T_m$(2)/dt [K/(10 years)] (c) and the corresponding parameters of the relationship d$T_{ms}$/d$T_{NHs}$, d$T_m$(1)/d$T_{NHs}$, d$T_m$(2)/d$T_{NHs}$ for four subintervals: 1960-1976, 1976-1992, 1992-2008, 2008-2024. The standard deviations are indicated in parentheses. The most statistically significant estimates are highlighted.

(a)

| Parameters | 1960−1976 | 1976−1992 | 1992−2008 | 2008−2024 |
|---|---|---|---|---|
| d$T_{ms}$/dt [K/(10 years)] | **−6.6(±1.9)** | **−5.2(±2.5)** | −0.4(±3.0) | −0.8(±1.8) |
| d$T_{ms}$/d$T_{NHs}$ | 10.4(±5.6) | 0.0(±6.2) | −0.2(±6.5) | −1.3(±2.6) |

(b)

| Parameters | 1960−1976 | 1976−1992 | 1992−2008 | 2008−2024 |
|---|---|---|---|---|
| d$T_m$(1)/dt [K/(10 years)] | **−4.9(±1.2)** | **−8.2(±1.5)** | 0.5(±1.6) | −2.7(±2.2) |
| d$T_m$(1)/d$T_{NHs}$ | 6.6(±3.9) | **−11.5(±5.0)** | 0.6(±3.3) | −3.9(±3.1) |

(c)

| Parameters | 1960−1976 | 1976−1992 | 1992−2008 | 2008−2024 |
|---|---|---|---|---|
| d$T_m$(2)/dt [K/(10 years)] | **−5.8(±1.5)** | **−7.5(±1.7)** | 0.9(±2.0) | −2.0(±1.8) |
| d$T_m$(2)/d$T_{NHs}$ | **10.6(±4.2)** | −8.3(±5.2) | 0.4(±4.3) | −2.4(±2.6) |

Similarly to [19], along with trends, an analysis was carried out of the relationship between temperature changes at the mesopause region and at the surface, in particular for the NH as a whole, $T_{NHs}$, based on data for the time interval 1960-2024 (Fig. 5, Tables 1, 2). In this case, the parameters d$T_{ms}$/d$T_{NHs}$, d$T_m$(1)/d$T_{NHs}$, d$T_m$(2)/d$T_{NHs}$ were estimated, characterizing the relationship of variations in $T_{ms}$, $T_m$(1), $T_m$(2) with variations in $T_{NHs}$ using the corresponding linear regressions. Linear regression coefficients $T_{ms}$, $T_m$(1), $T_m$(2) on $T_{NHs}$ can be used for estimation of the sensitivity parameters d$T_{ms}$/d$T_{NHs}$, d$T_m$(1)/d$T_{NHs}$, d$T_m$(2)/d$T_{NHs}$ of changes in $T_{ms}$, $T_m$(1), $T_m$(2) to changes in $T_{NHs}$. Quantitative estimates of the relationship parameters d$T_{ms}$/d$T_{NHs}$, d$T_m$(1)/d$T_{NHs}$, d$T_m$(2)/d$T_{NHs}$ are presented in Table 1 for the entire analyzed time interval 1960–2024 and for two 32-year subintervals (1960–1992 and 1992–2024), and for shorter 16-year subintervals in Table 2.

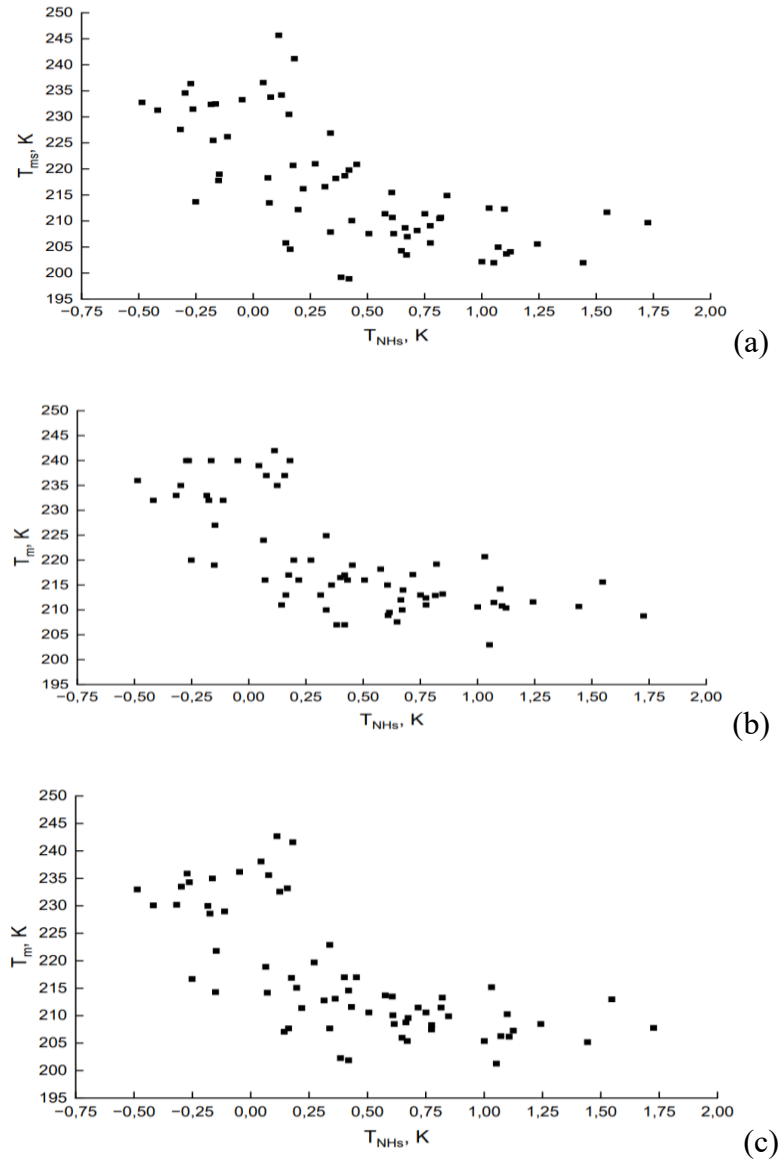

Fig. 5. Interannual variations in winter temperature at the mesopause level $T_{ms}$, $T_m(1)$, $T_m(2)$ according to measurements at the ZSS IAP RAS depending on the corresponding winter variations in the surface temperature of the Northern Hemisphere $T_{NHs}$ in 1960−2024.

According to Table 1 for the entire analyzed time interval, estimates of the coupling parameters $dT_m(1)/dT_{NHs}$, $dT_m(2)/dT_{NHs}$ and $dT_{ms}/dT_{NHs}$, with and without taking into account the contribution of solar radiation flux variations, are close: $-(15 \div 16)$. This indicates that the influence of solar radiation variability on trends is not significant at time intervals of the order of 6−7 decades. The corresponding estimates for two subintervals in Table 1a,b,c differ significantly, they are statistically significant for the first subinterval 1960−1992 and insignificant for the second subinterval (1992−2024). At the same time, the estimates of the parameters $dT_m(1)/dT_{NHs}$ and $dT_m(2)/dT_{NHs}$ with the exclusion of the solar activity influence are greater in absolute value than the estimate of $dT_{ms}/dT_{NHs}$ with the influence of solar activity.

According to Table 2 for shorter time subintervals, along with negative estimates $dT_m(1)/dT_{NHs}$ and $dT_m(2)/dT_{NHs}$, in particular for the subinterval 1976−1992, positive estimates $dT_m(1)/dT_{NHs}$ and $dT_m(2)/dT_{NHs}$, in particular for the 1960−1976 subinterval, were obtained. The results obtained are consistent with the conclusions made in [19] about significant differences, up to the sign, between the parameters of the relationship of temperature variations in the mesopause region and near the surface on longer time scales of several decades and on shorter shorter time scales (from interannual to interdecadal).

Abrupt change in temperature regime near mesopause in the 1970s also manifests itself when analyzing the connection between winter variations in $T_m$ based on 65 years of measurement data at the ZSS IAP RAS with the corresponding variations in the hemispheric surface temperature $T_{NHs}$ in Fig. 4. As a result of a sharp drop in $T_m$ in the 1970s there has been a transition to new temperature conditions characteristic of recent decades, starting from the 1980s. At the same time, as already noted in [19], mutual temperature changes in the mesopause region and near the surface differ significantly for interannual and interdecadal variations. The negative correlation between variations in $T_m$ and $T_{NHs}$, which is significant for longer-term variations, does not manifest itself in interannual variability. The manifestation of a positive correlation between $T_m$ and $T_{NHs}$ for individual subintervals indicates a difference in the mechanisms influencing on the relationship between temperature variations in the upper and lower atmosphere with characteristic scales of the order of a year and decades.

Analysis of distribution functions for winter $T_{ms}$ values for the entire analyzed period 1960–2024 (Fig. 6) indicates the manifestation of different temperature regimes in the mesopause region before and after the noted temperature jump in the 1970s. This was previously identified for a shorter time interval of 1960−2015 [19]. Distribution function for winter $T_{ms}$ values for the entire analyzed period 1960–2024 is characterized by two maxima: in the ranges of 230–235 K and 210–215 K, with a minimum between them in the range of 220–225 K (Fig. 6). The minimum of the distribution function corresponds to the years of temperature jump in the 1970s. The maximum with an average $T_{ms}$ value in the range of 230–235 K corresponds to the first time subinterval. A maximum in the range of 210–215 K characterizes $T_{ms}$ after the temperature jump in the 1970s and transition to a new regime.

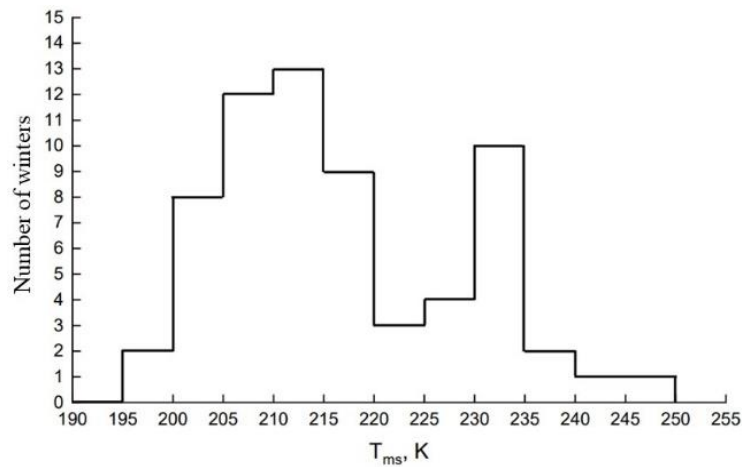

Fig. 6. Number of winters with temperature values in the mesopause region $T_{ms}$ in different ranges according to data for the entire period of measurements at the ZSS IAP RAS (1960–2024).

In [14,19], the synchronicity of the noted jump in the temperature regime in the mesopause region according to measurements at the ZSS IAP RAS with the well-known climate shift in the 1970s, noted in various near-surface climatic features, including the intensification of the Aleutian atmospheric center of action, was noted (see, for example, [19]). Climate shift 1976–1977 in the Pacific sector is associated with qualitative changes in the processes of formation of El Niño phenomena [25–27]. In particular, in [25], the manifestation of more frequent El Niño events and rarer La Niña events has been noted since the second half of the 1970s. A more significant connection at the end of the 20th century with the El Niño phenomena was also noted for the characteristics of the winter North Pacific centers of atmospheric action, the Aleutian minimum and the Hawaiian maximum [28] (see also [29, 30]). In [31], significant changes are noted in the frequency of various interphase transitions for El Niño phenomena (to which the strongest interannual variations in global surface temperature are associated) against the background of longer-period (with a period of several decades) variations and, in general, positive secular trend in global surface temperature. In [32, 33], a tendency is noted for the intensification and frequency of El Niño phenomena with global warming.

Figure 7 characterizes the cross-wavelet coherence of winter temperature variations in the mesopause region $T_{ms}$ according to data for the time interval 1960–2024 with El Niño indices: (a) Nino3, (b) Nino3.4, (c) Nino4. According to Fig. 7 in recent decades, the connection between winter temperature variations in the mesopause region and El Niño indices has been more significant for interdecadal and shorter-period variations.

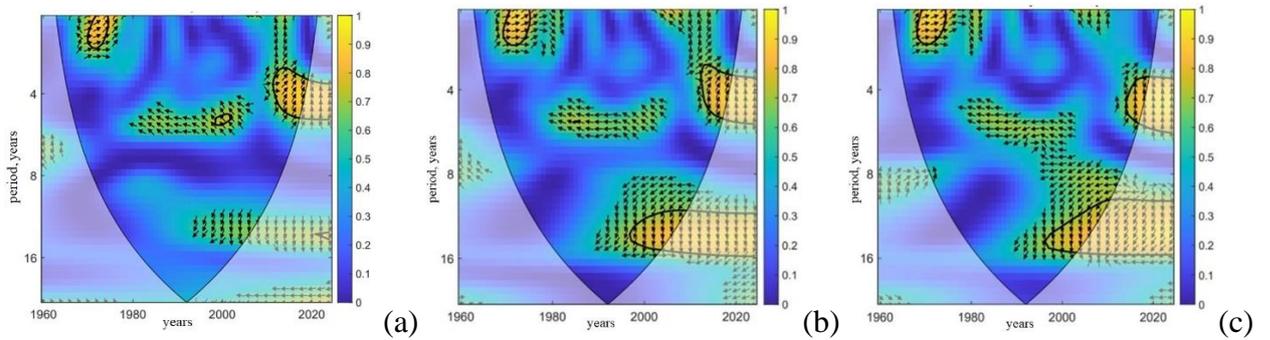

Fig. 7. Cross-wavelet coherence of winter temperature variations in the mesopause region $T_{ms}$ according to data for the period 1960–2024 with El Nino indices: (a) Nino3, (b) Nino3.4, (c) Nino4. Solid lines separate areas of edge effects, and thick lines limit areas of significant coherence (with coherence different from zero at the significance level $p = 0.05$). The arrows characterize the phase shift: arrows to the right mean in-phase, arrows to the left mean anti-phase.

As for long-term tendencies, no significant coherence of the longest-period variations in $T_{ms}$ and $T_m$ with $T_{NHs}$ was detected in [19] from observations for the time interval 1960−2015 with the use of cross-wavelet analysis. It was noted that although the series of data for variations in $T_{ms}$ and $T_m$ from measurements at the ZSS IAP RAS is the longest in the world (since 1959), its duration so far makes it possible to detect the coherence of variations in $T_{ms}$ and $T_m$ with $T_{NHs}$ only with characteristic times of less than two decades.

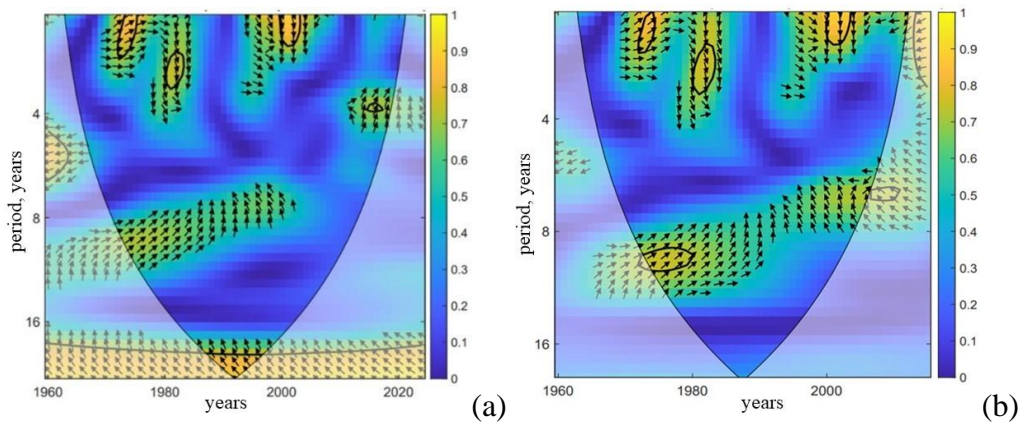

Fig. 8. Cross-wavelet coherence of winter temperature variations in the mesopause region $T_{ms}$ with variations in hemispheric surface temperature $T_{NHs}$ (a) according to data for the period 1960–2024 (a) and 1960−2015 (b). Solid lines delineate areas of edge effects, and thick lines delineate areas of significant coherence (with non-zero coherence at the $p = 0.05$ significance level). The arrows characterize the phase shift: arrows to the right mean in-phase, arrows to the left mean anti-phase.

Figure 8a, which characterizes the cross-wavelet coherence of winter temperature variations in the mesopause region $T_{ms}$ with variations of the hemispheric surface temperature $T_{NHs}$ according to 65-year data for 1960–2024, already reveals a significant coherence of their long-term (more than two decades) variations. The correlation of long-term variations is negative. For comparison, Fig. 8b presents the results of a similar analysis for 56 years of data (1960−2015), as in [19], without the manifestation of significant coherence of long-term variations of $T_{ms}$ and $T_{NHs}$.

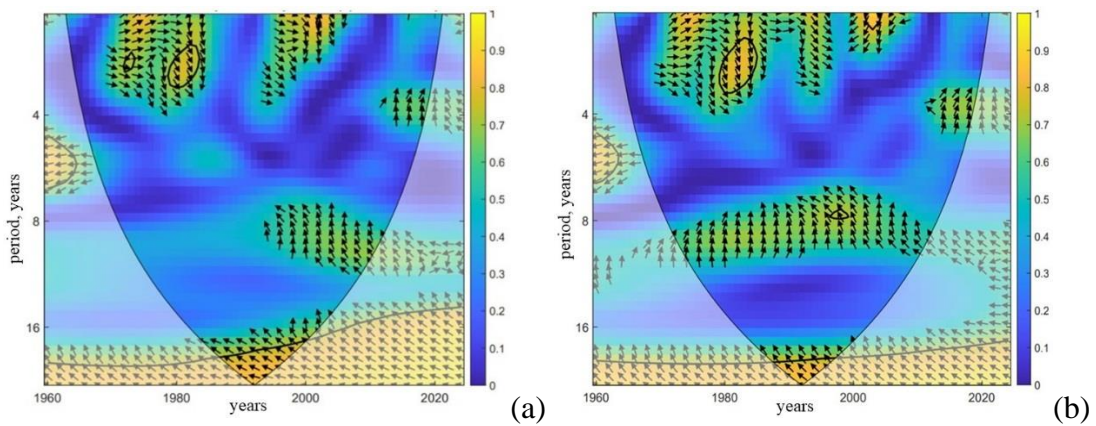

Fig. 9. Cross-wavelet coherence with variations in hemispheric surface temperature $T_{NHs}$ temperature in the mesopause region $T_m(1)$ (a) and $T_m(2)$ (b) according to winter data for the period 1960–2024. Solid lines delineate areas of edge effects, and thick lines delineate areas of significant coherence (with non-zero coherence at the $p = 0.05$ significance level). The arrows characterize the phase shift: arrows to the right mean in-phase, arrows to the left mean anti-phase.

Figure 9 characterizes cross-wavelet coherence with variations in hemispheric surface temperature $T_{NHs}$ with temperature in the mesopause region $T_m(1)$ (a) and $T_m(2)$ (b) according to winter data for the time interval 1960–2024. The results of the analysis of the relationship between $T_m(1)$ and $T_{NHs}$ indicate a more significant coherence of their long-term variations in recent years than for $T_{ms}$ and $T_{NHs}$. The results obtained confirm the presence of a significant negative correlation of long-term temperature variations in the mesopause region with variations in surface temperature, which was expected in accordance with model prognostic simulations [19].

**DISCUSSION AND CONCLUSION**

Results of measurements at the ZSS IAP RAS since the late 1950s indicate that, against the background of a general increase in surface temperature, the temperature at the mesopause level has decreased significantly, in particular in winter, with a sharp temperature shift in the 1970s and weakening cooling rates in recent decades. As part of the NDMC system, regular measurements of temperature in the mesopause region at other stations began to be carried out in the 1980s, i.e. after the sharp temperature changes in the mesopause region were noted in the 1970s according to measurements at the ZSS IAP RAS. Comparison of data obtained at the ZSS IAP RAS with available measurement data at other stations of the NDMC system indicates the general consistency of data for temperature in the mesopause region and its trend over the past decades (see, for example, [9,10,17]). Special comparisons in 2015 of temperature values in the mesopause region by measurements at the ZSS IAP RAS with those obtained from lidar measurements of temperature profiles in the Moscow region revealed their good agreement [22].

Multi-year increase in global surface temperature is accompanied by the temperature decrease in the mesopause region with a slowdown in recent years. At the same time, a sharp decrease in temperature in the mesopause region in the 1970s was revealed [14,19] synchronously with the well-known climate shift of the 1970s, previously identified in different climatic features near the surface. The noted climate change with the transition to a new regime is associated with changes in the El Niño phenomena. The results of cross-wavelet analysis using the results of measurements at the ZSS IAP RAS for the time interval 1960−2024 indicate a more significant connection between temperature variations in the mesopause region and El Niño indices in recent decades.

Significant coherence of the long-term temperature variations in the mesopause region with the hemispheric surface temperature from observational data for the time interval 1960−2024 was revealed with the use of the cross-wavelet-analysis. Previously, such a relationship was not noted in [19] based on the corresponding data for a shorter time interval. At the same time, it was revealed by longer-period climate model simulations for the 20−21st centuries, and it was assumed that to identify the corresponding significant coherence of long-term anti-phase variations at the mesopause level and at the surface according to observational data, longer measurements are necessary.

The noted features of temperature changes in the mesopause region against the background of changes in surface temperature are an important indicator for quantitative assessment of the comparative role of natural and anthropogenic factors in the formation of modern climate changes.

**REFERENCES**

1. Roble R.G., Dickinson R. E. How will changes in carbon dioxide and methane modify the mean structure of the mesosphere and thermosphere? *Geophys. Res. Lett.,* 1989, **16**, 1441–1444.
2. Golitsyn G.S., Semenov A. I., Shefov N. N. et al. Long-term temperature trends in the middle and upper atmosphere. *Geophys. Res. Lett.,* 1996, **23**, 1741–1744.
3. Mokhov I.I., Eliseev A.V. Tropospheric and stratospheric temperature cycle: Tendencies of change. *Izv., Atmos. Oceanic Phys.*, 1997, **33** (4), 415-426.
4. Semenov A.I., Shefov N.N., Givishvilli G.V., Leshchenko L.N., Lysenko E.V., Rusina V.Ya., Fishkova L.M., Martsvaladze N.M., Toroshelidze T.I., Kashcheev B.L., Oleynikov A.N. Seasonal pequalirities of long-term temperature trends of the middle atmosphere. *Doklady Earth Sci.*, 2000, **375**, 1286-1289.
5. Semenov A.I., Shefov N. N., Lysenko E. N. et al. The season peculiarities of behavior of the long-term temperature trends in the middle atmosphere on the mid-latitudes. Phys. Chem. Earth., 2002, **27**, 529–534.

6. Beig G., Keckhut P., Lowe R. P. et al. Review of mesospheric temperature trends // Rev. Geophys., 2003, **41** (4), 1015, doi:10.1029/2002RG000121.
7. Khomich V.Y., Semenov A.I., Shefov N.N. Airglow as an Indicator of Upper Atmospheric Structure and Dynamics. Berlin, Springer, 2008.
8. Climate Change 2007: Physical Science Basis. Contribution of Working Group I to the Fifth Assessment Report of the Intergovernmental Panel on Climate Change / Eds: Solomon S., Qin D., Manning M. et al. Cambridge: Cambridge Univ. Press, 2007.
9. Beig G. Long-term trends in the temperature of the mesosphere/lower thermosphere region: 1. Anthropogenic influences. *J. Geophys. Res.*, 2011, **116**, A00H11, doi:10.1029/2011JA016646.
10. Beig G. Long-term trends in the temperature of the mesosphere/lower thermosphere region: 2. Solar response. *J. Geophys. Res.*, 2011, **116**, A00H12, doi:10.1029/2011JA016766.
11. Qian L., Lastovicka J., Roble R. G. et al. Progress in observations and simulations of global change in the upper atmosphere. *J. Geophys. Res.*, 2011, **116**, A00H03, doi:10.1029/2010JA016317
12. Bindoff N.L., Stott P. A., AchutaRao K.M. et al. Detection and attribution of climate change: from global to regional / In: Climate Change 2013: Physical Science Basis. Contribution of Working Group I to the Fifth Assessment Report of the Intergovernmental Panel on Climate Change / Eds: Stocker T. F., Qin D., Plattner G.-K. et al. Cambridge: Cambridge Univ. Press, 2013, 867–952.
13. Lübken F.-J., Berger U., Baumgarten U. Temperature trends in the midlatitude summer mesosphere. *J. Geophys Res Atmos.*, 2013, **118**, 13347–13360. doi:1002/2013JD020576
14. Mokhov I.I., Semenov A.I. Nonlinear temperature changes in the atmospheric region of the atmosphere against the background of global climate changes. *Doklady Earth Sci.*, 2014, **456** (2), 741-744.
15. Laštovička J., Beig G., Marsh D.R. Response of the mesosphere-thermosphere-ionosphere system to global change – CAWSES-II contribution. Progr. Earth Planet. Sci., 2014, **1** (21), 1-19.
16. Perminov V.I., Semenov A.I., Medvedeva I.V., Zheleznov Yu.A. Variability of mesopause temperature from the hydroxyl airglow observations over midlatitudinal sites, Zvenigorod and Tory, Russia. *Adv. Space Res.*, 2014, **54**, 2511–2517.
17. She C.-Y., Krueger D. A., Yuan T. Long-term midlatitude mesopause region temperature trend deduced from quarter century (1990–2014) Na lidar observations. *AnGeo Comm.*, 2015, **33**, 363–369.
18. Kalicinsky C. et al. Long-term dynamics of OH∗ temperatures over central Europe: trends and solar correlations. *Atmos. Chem. Phys.*, 2016, **16**, 15033–15047.
19. Mokhov I.I., Semenov A.I., Volodin E.M., Dembitskaya M.A. Changes of cooling near mesopause under global warming from observations and model simulations. *Izv., Atmos. Oceanic Phys.*, 2017, **53** (4), 383-391.
20. Santer B.D. et al. Contributions of anthropogenic and natural forcing to recent tropopause height changes. *Science*, 2003, **301**, 479–483.
21. Assessment Report on Climate Change and its Consequences on the Territory of the Russian Federation. V. I. Climate change. Moscow, Roshydromet, 2008. 227 pp. (in Russian)
22. Ivanov M.S., Mokhov I.I., Semenov A.I., Sumarokov V.V. Comparison of temperature in the upper mesosphere from lidar measurements, satellite and model data and from ground-based measurements of hydroxyl emission. *Research Activities in Atmospheric and Oceanic Modelling*, E. Astakhova (ed.), 2017, WCRP Rep. No. 12/2017, S. 2, 13–14.
23. Dalin P., Perminov V., Pertsev N., Romejko V. Updated long-term trends in mesopause temperature, airglow emissions, and noctilucent clouds, *J. Geophys. Res.*, 2020, **125**, e2019JD030814. doi: 10.1029/2019JD030814.
24. Jevrejeva S., Moore J. C., Grinsted A. Influence of the Arctic Oscillation and El Nino-Southern Oscillation (ENSO) on the conditions in the Baltic Sea: The wavelet approach. *J. Geophys. Res.*, 2003, **108** (D21), 4677, doi:10.1029/2003JD003417.
25. Trenberth K.E., Hoar T. J. El Nino and climate change. *Geophys. Res. Lett.*, 1997, **24** (23), 3057–3060.
26. Zhang Y., Wallace J. M., Battisti D. S. ENSO-like interdecadal variability: 1900–93. J. Climate, 1997, **10** (5), 1004–1020.
27. Meehl G.A., Hu A., Santer B.D. The mid-1970s climate shift in the Pacific and the relative roles of forced versus inherent decadal variability. *J. Climate*, 2009, **22**, 780–792.
28. Mokhov I.I., Khon V.Ch. Interannual variability and long-term tendencies of change in atmospheric centers of action in the Northern Hemisphere: Analyses of observational data. *Izv., Atmos. Oceanic Phys.*, 2005, **41** (6), 657-666.
29. Mokhov I.I., Chernokulsky A.V., Osipov A.M. Atmospheric centers of action in the Northern and Southern Hemispheres: Features and variability. Russ. Meteorol. Hydrol., 2020, **45** (11), 749-761.
30. Intense Atmospheric Vortices and their Dynamics. I.I. Mokhov, M.V. Kurgansky, O.G. Chkhetiani (eds.). Moscow, GEOS, 2018. 482 pp. (in Russian)
31. Mokhov I.I. Changes in the frequency of phase transitions of different types of El Nino phenomena in recent decades. *Izv., Atmos. Oceanic Phys.*, 2022, **58** (1), 1-6.
32. Mokhov I.I., Eliseev A.V., Khvorostyanov D.V. Evolution of the characteristics of interannual climate variability associated with the El Nino and La Nina phenomena. *Izv., Atmos. Oceanic Phys.*, 2000, **36** (6), 681-690.
33. Mokhov I.I., Khvorostyanov D. V., Eliseev A. V. Decadal and longer term changes in El Nino – Southern Oscillation characteristics. *Intern. J. Climatol.*, 2004, **24**, 401–414.